\documentclass{article}
\usepackage{amssymb,amsmath,latexsym,amscd,amsfonts}
\usepackage{graphics}
\usepackage{epsfig}

\newtheorem{theorem}{Theorem}[section]

\newtheorem{definition}{Definition}[section]

\newtheorem{lemma}{Lemma}[section]
\newtheorem{conjecture}{Conjecture}[section]

\newtheorem{corollary}{Corollary}[section]

\def\hpic #1 #2 {\mbox{$\begin{array}[c]{l}
\epsfig{file=#1,height=#2} \end{array}$}}
\def\vpic #1 #2 {\mbox{$\begin{array}[c]{l}
\epsfig{file=#1,width=#2}\end{array}$}}

\begin{document}

\title{$\mathcal{C}_3$, Semi-Clifford and Generalized Semi-Clifford Operations}
\author{Salman Beigi\thanks{Institute for Quantum Information, California Institute of Technology, Pasadena, CA} \hspace{.7in}
 Peter W. Shor\thanks{Department of Mathematics, Massachusetts Institute of Technology, Cambridge, MA  } }
\date{}
\maketitle

\begin{abstract}
Fault-tolerant quantum computation is a basic problem in quantum
computation, and teleportation is one of the main techniques in
this theory. Using teleportation on stabilizer codes, the most
well-known quantum codes, Pauli gates and Clifford operators
can be applied fault-tolerantly. Indeed, this technique can be
generalized for an extended set of gates, the so called
${\mathcal{C}}_k$ hierarchy gates, introduced by Gottesman and
Chuang (Nature, 402, 390-392). ${\mathcal{C}}_k$ gates are a
generalization of Clifford operators, but our knowledge of these
sets is not as rich as our knowledge of Clifford gates. Zeng et
al. in (Phys. Rev. A 77, 042313) raise the question of the
relation between ${\mathcal{C}}_k$ hierarchy and the set of
semi-Clifford and generalized semi-Clifford operators. They
conjecture that any ${\mathcal{C}}_k$ gate is a generalized semi-Clifford
operator. In this paper, we prove this conjecture for $k=3$. Using
the techniques that we develop, we obtain more insight on how to
characterize ${\mathcal{C}}_3$ gates. Indeed, the more we understand
${\mathcal{C}}_3$, the more intuition we have on ${\mathcal{C}}_k$,
$k\geq 4$, and then we have a way of attacking the conjecture for
larger $k$.
\end{abstract}

\section{Introduction}	      
The theory of fault-tolerant quantum computation is one of the
main parts of the theory of quantum computation. In this theory we
are introducing a quantum code, a universal set of gates, and
then a method to apply these gates fault-tolerantly
\cite{chuang}. The most important quantum codes are quantum
stabilizer codes, and teleportation is an idea to apply a
universal set of quantum gates on these codes \cite{nature}.

It is well-known that all Pauli gates as well as Clifford
operators can be applied fault-tolerantly using teleportation.
However, these are not the only gates with such a property.
Indeed, a Clifford operator can be applied fault-tolerantly via
teleportation because by conjugation it sends the Pauli group to
itself. Generalizing this idea, we can define the so called
${\mathcal{C}}_k$ hierarchy operators.

\begin{definition}\label{def:ck}
Let ${\mathcal{P}}={\mathcal{C}}_{1}$ denote the Pauli group. For $k\geq 1$ define
\begin{equation}{\mathcal{C}}_{k+1}=\{U:\ U{\mathcal{P}}U^\dagger \subseteq {\mathcal{C}}_k \}.
\end{equation}
\end{definition}

Gottesman and Chuang in \cite{nature} introduce these sets and
show that all ${\mathcal{C}}_k$ gates can be applied
fault-tolerantly via teleportation. However, our knowledge of these
operators is poor.

By definition, ${\mathcal{C}}_2$ is the Clifford group, and there is
a rich theory for characterizing and representing these operators
\cite{clifford}. Also, by definition ${\mathcal{C}}_k\subseteq
{\mathcal{C}}_{k+1}$. But for
$k\geq 3$, ${\mathcal{C}}_k$ is no longer a group \cite{bei}, and because of that, the problem of
studying ${\mathcal{C}}_k$ hierarchy turns out to be a hard one.
However, if we restrict ourselves to diagonal gates in
${\mathcal{C}}_k$, it is a group \cite{bei}.

Let us think of the diagonal gates in another direction. If $U\in
{\mathcal{C}}_k$ is diagonal, then $U$ commutes with all $\sigma_z$
operators, and then $U{\mathcal{P}}U^\dagger$ contains a maximal
abelian subgroup of ${\mathcal{P}}$. Also, for such a $U$, it is not
hard to see that $Q_1UQ_2$ is in ${\mathcal{C}}_k$ and satisfies
the same property for all Clifford operators $Q_1$ and $Q_2$. This
observation leads us to the following definition.

\begin{definition}\label{def:semi-clifford}
A unitary operator $U$ acting on $n$ qubits is called
semi-Clifford if $U{\mathcal{P}}U^{\dagger}$ contains a maximal
abelian subgroup of ${\mathcal{P}}$. In other words, $U$, under
conjugation, sends a maximal abelian subgroup of ${\mathcal{P}}$ to
another maximal abelian subgroup of ${\mathcal{P}}$.
\end{definition}

Semi-Clifford operators are first defined in \cite{semi-clifford}
for one qubit operators, and are generalized for $n$ qubits in
\cite{bei}. Also in \cite{bei}, Zeng et al. raise the question
of the relation between semi-Clifford operators and
${\mathcal{C}}_k$ hierarchy. They show that all ${\mathcal{C}}_k$
hierarchy gates are semi-Clifford if $n=1,2$ ($n$ is the number
of qubits). For $n=3$ they prove the same property if $k=3$, and
propose the following conjecture for larger $n$.

\begin{conjecture}\label{conj1} {\rm \cite{bei}}
All gates in ${\mathcal{C}}_3$ are semi-Clifford operations.
\end{conjecture}

Moreover, for $k\geq 4$ by giving an example they show that there
are gates in ${\mathcal{C}}_k$ which are not semi-Clifford. But they
realize that those gates are {\it generalized semi-Clifford}.

\begin{definition} A generalized semi-Clifford operator on $n$ qubits
is defined to send, by conjugation, the linear span of at least
one maximal abelian subgroup of ${\mathcal{P}}$ to the linear span
of another maximal abelian subgroup of ${\mathcal{P}}$.
\end{definition}

Clearly, any semi-Clifford operator is generalized
semi-Clifford. But we can think of an abelian group of $2^n$
diagonal matrices which are all linearly independent and are
different from $\sigma_z$ gates; then, the span of this group is
the same as the span of all $\sigma_z$ operators in ${\mathcal{P}}$.
Thus, semi-Clifford and generalized semi-Clifford operators are not
the same.

Here is the second conjecture made in \cite{bei}.

\begin{conjecture} \label{conj2} {\rm \cite{bei}} All gates in ${\mathcal{C}}_k$ are generalized
semi-Clifford operations.
\end{conjecture}

The main result of this paper is that Conjecture \ref{conj2} holds for $k=3$.

\noindent\begin{theorem}\label{thm:main-result} Every gate in ${\mathcal{C}}_3$ is generalized semi-Clifford.
\end{theorem}

\subsection{Related works}

\noindent
Gottesman and Mochon (personal communication) have disproved Conjecture \ref{conj1}, i.e.
they have found a gate in ${\mathcal{C}}_3$ which is not semi-Clifford.
Here we briefly discuss their counterexample.

Consider seven qubits and call them $A_1, A_2, A_3, B_1, B_2, B_2$, and $R$. Let $U$ be the multiplication of the three controlled-swap gates which act on $(R, A_i, B_i)$, $i=1,2, 3$ (it swaps $A_i$ and $B_i$ if $R$ is $\vert 1\rangle$). Also, let $V$ be the multiplication of four controlled-$\sigma_z$ gates which act on $(A_1, A_2, A_3)$, $(A_1, B_2, B_3)$, $(B_1, A_2, B_3)$, and $(B_1, B_2, A_3)$. By computing the action of $UV$ on Pauli matrices it can be seen that $UV\in {\mathcal{C}}_3$; however, $VU$ is not in ${\mathcal{C}}_3$ since $(VU) \sigma_x (VU)^{\dagger}$, where $\sigma_x$ acts on qubit $R$, does not belong to ${\mathcal{C}}_2$.

Now, we claim that $UV$ is a ${\mathcal{C}}_3$ gate which is not semi-Clifford. Suppose $UV$ is semi-Clifford; then, by Proposition 1 of \cite{bei} there are Clifford operations $Q_1, Q_2$ such that $D=Q_1UVQ_2$ is diagonal. On the other hand, since $UV$ is in ${\mathcal{C}}_3$, $D\in {\mathcal{C}}_3$, which means that $D{\mathcal{P}}D^{\dagger} \subseteq {\mathcal{C}}_2$. Note that, for any $\sigma\in {\mathcal{P}}$ and $Q\in {\mathcal{C}}_2$, $\bar{\sigma}$ and $\bar{Q}$ (the entry-wise complex conjugate of $\sigma$ and $Q$) also belong to ${\mathcal{P}}$ and ${\mathcal{C}}_2$, respectively. Hence, $D{\mathcal{P}}D^{\dagger} \subseteq {\mathcal{C}}_2$ implies $\bar{D}{\mathcal{P}}D^T \subseteq {\mathcal{C}}_2$, or equivalently $\bar{D}=D^{\dagger} = Q_2^{\dagger}VUQ^{\dagger}_1\in {\mathcal{C}}_3$. Therefore, $VU$ is in ${\mathcal{C}}_3$, which is a contradiction.

\subsection{Structure of the paper}

\noindent
In Section 2, we fix some notations on Pauli
operators and then review the characterization of Clifford
operations from \cite{clifford}, which represents each Clifford
gate by a $C$-matrix (a symplectic matrix) and an $h$-vector over the binary field.

In Section 3, we formulate our main idea for proving
Theorem \ref{thm:main-result}, which is to express the whole
assumptions in terms of some relations on $C$-matrices and
$h$-vectors.

In Section 4, we reduce the problem to a special
case where $C$-matrices contain a block of zeros, which is easier to
handle.

In Section 5, we are trying to obtain more information
from the representation of Clifford operations based on
$C$-matrices and $h$-vectors. Indeed, $C$ and $h$
characterize Clifford operators up to an overall phase. Thus,
having $C$ and $h$, we should be able to determine entries of a
Clifford operator as a matrix, up to an overall phase. In Section 5, we explicitly find the matrix
representation of Clifford operations in the special case
introduced in Section 4. The generalization of
these results are studies in Appendix A, in which each Clifford operator is expressed as
a linear combination of Pauli matrices, and the coefficients of this expansion are computed.

In Section 6, we find a formula that given two Clifford operators
$Q$ and $Q'$ represented by $(C, h)$ and $(C', h')$, respectively, expresses whether $QQ'=Q'Q$ or $QQ'=-Q'Q$. Notice that this is a valid question since the representation of Clifford
operations by $C$-matrices and $h$-vectors, is independent of an overall
phase, and cannot distinguish $Q'Q$ and $-Q'Q$.

In Section 7, we put all pieces together and finish
the proof of Theorem \ref{thm:main-result}.

\section{Preliminaries} \label{sec:prem}

\subsection{Pauli operators}

\noindent
First of all let us fix some notations for Pauli matrices.
\begin{equation}\sigma_{00}=\tau_{00}=\sigma_0=
\left(
\begin{array}{cc}
    2 & 2 \\
    2 & 2 \\
  \end{array}
\right)
, \hspace{.5in} \sigma_{01}=\tau_{01}=\sigma_x=\left(
                                                 \begin{array}{cc}
                                                   0 & 1 \\
                                                   1 & 0 \\
                                                 \end{array}
                                               \right)
,\end{equation}
\begin{equation}\sigma_{10}=\tau_{10}=\sigma_z=\left(\begin{array}{cc}
  1 & 0 \\
  0 & -1
\end{array}\right), \hspace{.5in} i\sigma_{11}=\tau_{11}=i\sigma_y=\left( \begin{array}{cc}
  0 & 1 \\
  -1 & 0
\end{array} \right).\end{equation}
Pauli operators over $n$ qubits are denoted by $\sigma_a$ and
$\tau_a$, where $a=\left( \begin{array}{c}
  v \\
  w
\end{array} \right)$ is in ${\mathbf{Z}}^{2n}_2$ and
\begin{equation}\sigma_a=\sigma_{v_1w_1}\otimes \cdots \otimes \sigma_{v_nw_n},\end{equation}
\begin{equation}\tau_a=\tau_{v_1w_1}\otimes \cdots \otimes \tau_{v_nw_n}.\end{equation}
Also, we represent the phases $\pm 1, \pm i$ by $i^\delta
(-1)^\epsilon$, where $\delta, \epsilon\in {\mathbf{Z}}_2$. Then, it
is easy to see that the multiplication of two Pauli operators
$\big(i^{\delta_1}(-1)^{\epsilon_1}\tau_{a_1}\big)
\big(i^{\delta_2}(-1)^{\epsilon_2}\tau_{a_2}\big)$ is equal to
$i^{\delta}(-1)^{\epsilon}\tau_{a}$, where
\begin{eqnarray}\label{eq:pm}
\delta   & = &\delta_1+\delta_2,\\
\epsilon & =
&\epsilon_1+\epsilon_2+\delta_1\delta_2+a_2^TJ\,a_1,\\
a        & = &a_1+a_2,\label{eq:pm3}
\end{eqnarray}
in which $T$ denotes the transposed matrix,
\begin{equation}\label{eq:j}
J=\left( \begin{array}{cc}
  0 & I_n \\
  0 & 0
\end{array} \right),
\end{equation}
and $I_n$ is the identity matrix of size $n$. As a result,
\begin{equation}\label{eq:com}
\tau_a\tau_b=(-1)^{b^T P\, a} \tau_b\,\tau_a,
\end{equation}
where
\begin{equation}\label{eq:p}
P=J+J^T=\left( \begin{array}{cc}
  0 & I_n \\
  I_n & 0
\end{array} \right).
\end{equation}

\subsection{Clifford operators}

\noindent
Assume that $Q$ is a Clifford operator. By definition for each
Pauli matrix $\tau_a$, $Q\tau_aQ^\dagger$ is also a Pauli
operator. According to Eq. (\ref{eq:pm3}) and
\begin{equation}\label{eq:conju2}Q\tau_{a_1}\tau_{a_2}Q^\dagger=(Q\tau_{a_1}Q^\dagger)(Q\tau_{a_2}Q^\dagger)
\end{equation} to
compute the image of Pauli operators under conjugation by $Q$, it is
sufficient to know $Q\tau_{e_j}Q^\dagger$, $j=1,\dots, 2n$, where
$\{e_1,\dots , e_{2n}\}$ is the standard basis of
${\mathbf{Z}}^{2n}_2$ (all coordinates of $e_j$ are $0$ except the
$j$-th which is $1$). Hence, assume that
$Q\tau_{e_{j}}Q^\dagger=i^{d_j}(-1)^{h_j}\tau_{c_j}$, where
$d,h\in {\mathbf{Z}}^{2n}_2$, and $c_j\in {\mathbf{Z}}^{2n}_2$. Notice that
$\tau_{e_j}$ and then $Q\tau_{e_j}Q^\dagger$ are
hermitian, so $d_j$ can be determined in terms of $c_j$;
$d_j=c_j^TJ\,c_j$. Thus, if we define a $2n\times 2n$ matrix $C$ whose $j$-th column is equal to $c_j$, then
\begin{equation}\label{eq:d}
d=diag(C^TJ\,C),
\end{equation}
where $diag(M)$ denotes a vector whose $j$-th coordinate is the
$j$-th entry on the diagonal of $M$.

Now, using the matrix $C$ and Eq. (\ref{eq:conju2}) we can compute
$Q(i^{\delta_1}(-1)^{\epsilon_1}\tau_{a_1})Q^\dagger$ in terms of
$C$ and $h$. In fact,
$Q(i^{\delta_1}(-1)^{\epsilon_1}\tau_{a_1})Q^\dagger=i^{\delta_2}(-1)^{\epsilon_2}\tau_{a_2}$
where
\begin{eqnarray}\label{eq:conj1}
a_2         & = & Ca_1,\\
\delta_2    & = & \delta_1+d^Ta_1,\\
\epsilon_2  & = & \epsilon_1 + h^Ta_1 + a_1^T
lows(C^TJ\,C+dd^T)a_1 +\delta_1d^Ta_1,\label{eq:conj2}
\end{eqnarray}
in which $lows(M)$ denotes the strictly lower triangular part of
matrix $M$.

In order to determine the conditions that $C$ and $h$ should
satisfy, note that if $\tau_a$ and $\tau_b$ commute, then their
image under conjugation by $Q$, commute as well. Therefore, by
Eqs. (\ref{eq:com}) and (\ref{eq:conj1}), the map
$a\rightarrow Ca$ should preserve the {\it symplectic inner
product}, i.e. $a^TC^T P C b= a^TP b$, or equivalently
\begin{equation}\label{eq:symp}
C^TPC=P.
\end{equation}
Matrices $C$ that satisfy Eq. (\ref{eq:symp}) are called {\it
symplectic}.

Dehaene and De Moor prove the following theorem in
\cite{clifford}.

\begin{theorem} {\rm \cite{clifford}} \label{thm:cr} For any symplectic matrix $C$
(satisfying Eq. (\ref{eq:symp})) and any vector $h$ over ${\mathbf{Z}}_2$, there is a
unique (up to a phase) Clifford operator $Q$ such that
$Q(i^{\delta_1}(-1)^{\epsilon_1}\tau_{a_1})Q^\dagger=i^{\delta_2}(-1)^{\epsilon_2}\tau_{a_2}$
where $a_2, \delta_2$ and $\epsilon_2$ are given by Eqs.
(\ref{eq:conj1})-(\ref{eq:conj2}).
\end{theorem}

According to this theorem any Clifford operator $Q$ can be represented by
a pair $(C, h)$, where $C$ is a symplectic matrix. Then, to get a full
representation of Clifford operators as a group, it is sufficient
to compute the inverse and product of these operators based on
$C$-matrices and $h$-vectors.

\begin{theorem} {\rm \cite{clifford}} \label{thm:cmi}

\begin{enumerate}
\item[{\rm (a)}] Given $(C_1, h_1)$ and $(C_2, h_2)$ defining two Clifford
operators $Q_1$ and $Q_2$, respectively, the product
$Q_{12}=Q_2Q_1$ is represented by $(C_{12}, h_{12})$ such that
\begin{eqnarray}\label{eq:pc}
C_{12}            & = & C_2C_1,\\
h_{12} & = & h_1+C^T_1h_2+
diag(C^T_1lows(C^T_2J\,C_2+d_2d^T_2)C_1+ d_1d^T_2C_1),
\end{eqnarray}
where $d_1$ and $d_2$ are defined in Eq. (\ref{eq:d}).

\item[{\rm (b)}] Given $(C,h)$ defining a Clifford operator $Q$, the
inverse $Q'=Q^{-1}$ is represented by $(C', h')$ such that
\begin{eqnarray}\label{eq:ic}
C'             & = & C^{-1},\\
h' & = & C^{-T}h + diag(C^{-T}lows(C^TJ\,C+dd^T)C^{-1}+
d'd^TC^{-1}),
\end{eqnarray}
where $M^{-T}=(M^{-1})^T$, and $d'=diag(C^{-T}J\,C^{-1})$.
\end{enumerate}
\end{theorem}

Theorems \ref{thm:cr} and \ref{thm:cmi} give a full representation
of Clifford group. However, this representation is up to a phase, i.e.
two Clifford operations that differ only on a global
phase have the same $C$-matrix and $h$-vector.

As the last remark on this representation notice that each Pauli
operator is a Clifford gate as well. So, we can represent it
by a $C$-matrix and an $h$-vector. Also, by Eq. (\ref{eq:com})
every two Pauli matrices either commute or anti-commute.
Therefore, the $C$-matrix for all Pauli operators is identity.
More explicitly, the Pauli operator $\tau_a$ corresponds to
$(C=I_{2n}, h=Pa)$.

\section{Main ideas}\label{sec:main}

\noindent
In this section, we formulate our main idea for proving Theorem
\ref{thm:main-result}. Let $U$ be a $ {\mathcal{C}}_3$ gate on $n$
qubits. Then, by definition $U{\mathcal{P}}\,U^\dagger$ is a subset
of Clifford group, so if we define $Q_i=U \sigma_{e_i}
U^\dagger$, $i=1,\dots, 2n$, then $Q_i$ is a Clifford gate and
\begin{eqnarray}\label{eq:pauli1}
Q_i^2  & = & I,\\
Q_i\,Q_j & = &(-1)^{\delta_{i+n,j}}\,Q_j\,Q_i,\label{eq:pauli2}
\end{eqnarray}
where $\delta_{i+n,j}$ is the Kronecker delta function, and $i\leq j$.

Conversely, let $Q_1, \dots, Q_{2n}$ be Clifford operators that satisfy Eqs.
(\ref{eq:pauli1}) and (\ref{eq:pauli2}). Then, $Q_1, \dots, Q_{n}$
commute, and they have a common eigenvector $\vert
\alpha\rangle$ with eigenvalues $+1$ or $-1$. Let $Q_i\vert
\alpha\rangle=(-1)^{\lambda_i}\vert \alpha\rangle$, for $i=1, \dots, n$, where
$\lambda_i\in {\mathbf{Z}}_2$. Define the linear operator $U$ by
\begin{equation}U\vert x_1\dots x_{n}\rangle = Q_{n+1}^{x_1+\lambda_1}\dots Q_{2n}^{x_n+\lambda_n}\vert \alpha\rangle,\end{equation}
for every standard basis vector $\vert x_1\dots x_n\rangle$, where $x_1, \dots, x_n\in {\mathbf{Z}}_2$.

We claim that $U$ is unitary and $Q_i=U\sigma_{e_i}U^{\dagger}$ for every $i$. Since all $Q_i$'s are unitary, the vectors $U\vert x_1\dots x_{n}\rangle$ are normal. Also, if $(x_1, \dots, x_n)$ and $(y_1, \dots, y_n)$ are different, say at the first coordinate ($x_1+y_1=1$), then
\begin{eqnarray}
\langle x_1 \dots x_n\vert U^{\dagger}U\vert y_1\dots y_n\rangle & = & \langle \alpha \vert Q_{n+1}^{x_1+y_1}\dots Q_{2n}^{x_n+y_n}\vert \alpha\rangle \nonumber\\
& = & (-1)^{\lambda_1}\langle \alpha \vert Q_1 Q_{n+1}^{x_1+y_1}\dots Q_{2n}^{x_n+y_n}\vert \alpha\rangle \nonumber\\
& = & -(-1)^{\lambda_1}\langle \alpha \vert  Q_{n+1}^{x_1+y_1}\dots Q_{2n}^{x_n+y_n} Q_1\vert \alpha\rangle \nonumber\\
& = & -\langle \alpha \vert  Q_{n+1}^{x_1+y_1}\dots Q_{2n}^{x_n+y_n}\vert \alpha\rangle \nonumber\\
& = & -\langle x_1 \dots x_n\vert U^{\dagger}U\vert y_1\dots y_n\rangle.
\end{eqnarray}
Thus, $\langle x_1 \dots x_n\vert U^{\dagger}U\vert y_1\dots y_n\rangle=0$ and $U$ is unitary. $Q_i=U\sigma_{e_i}U^{\dagger}$ can be proved by showing that the action of $U^{\dagger}Q_iU$ on the basis vectors is equal to the action of $\sigma_{e_i}$.

This observation shows that any $C_3$ gate corresponds to a subgroup of the Clifford group which is isomorphism to the Pauli group, and vice versa. Also, it suggests to study subgroups of Clifford group, isomorphic to Pauli group,
instead of ${\mathcal{C}}_3$ gates directly.

\begin{theorem}\label{thm:main} Theorem \ref{thm:main-result} is
equivalent to the following:

Let ${\mathcal{G}}$ be a subgroup of the Clifford group which is isomorphic to the Pauli group. Then there exist maximal abelian subgroups ${\mathcal{H}}\subset {\mathcal{G}}$ and ${\mathcal{H'}}\subset {\mathcal{P}}$ such that the linear span of ${\mathcal{H}}$ is equal to the linear span of ${\mathcal{H'}}$. $\Box$
\end{theorem}

To proceed through this idea, let the group ${\mathcal{G}}$ be generated by Clifford operations $Q_1,
\dots, Q_{2n}$ that satisfy Eqs. (\ref{eq:pauli1}) and
(\ref{eq:pauli2}). As a Clifford operator, let $Q_i$ be represented
by the pair $(C_i, h_i)$, $i=1,\dots ,2n$. Then according to
Theorem \ref{thm:cmi}, Eq. (\ref{eq:pauli1}) implies
\begin{equation}\label{eq:pauli'1}
C_i^2  =  I,
\end{equation}
\begin{equation}
 h_i+C^T_ih_i+ diag(C^T_ilows(C^T_iJ\,C_i+d_id^T_i)C_i+
d_id^T_iC_i)  =  0,\label{eq:pauli'2}
\end{equation}
and Eq. (\ref{eq:pauli2}) implies
\begin{equation}\label{eq:pauli'3}
C_iC_j   =  C_jC_i,
\end{equation}
$$h_i+C^T_ih_j+
diag(C^T_ilows(C^T_jJ\,C_j+d_jd^T_j)C_i+ d_id^T_jC_i) \hspace{2in}
$$
\begin{equation}\label{eq:pauli'4}
 \hspace{1in}=
h_j+C^T_jh_i+ diag(C^T_jlows(C^T_iJ\,C_i+d_id^T_i)C_j+
d_jd^T_iC_j),
\end{equation}
for every $i,j$.

Notice that representing a Clifford operator by a
$C$-matrix and an $h$-vector is independent of a global phase. This is why here we
do not see the sign of Eq. (\ref{eq:pauli2}). Indeed, we need another equation, which we call {\it the sign
formula}, to compute this sign and get a full representation of
Eqs. (\ref{eq:pauli1}) and (\ref{eq:pauli2}).


Unlike Eqs. (\ref{eq:pauli'1}) and (\ref{eq:pauli'3}), Eqs. (\ref{eq:pauli'2}) and
(\ref{eq:pauli'4}) are not easy to handle, so we need to somehow reduce these equations
to simpler ones.

Let $Q$ be an arbitrary Clifford operator represented by the pair
$(C, h)$. Define $Q'_i=QQ_iQ^\dagger$; then, $Q'_1,\dots ,Q'_{2n}$
satisfy Eqs. (\ref{eq:pauli1}) and (\ref{eq:pauli2}) as well, so
they generate a subgroup of the Clifford group isomorphic to the
Pauli group. Also, since $Q$ sends Pauli operators to Pauli
operators under conjugation, proving the claim in Theorem \ref{thm:main} for
the group generated by $\{Q_i':\, i=1, \dots, 2n\}$, implies the theorem for the group
generated by $\{Q_i:\, i=1, \dots, 2n\}$.

This argument shows that in Eqs. (\ref{eq:pauli'1})-(\ref{eq:pauli'4}), we can replace the
symplectic matrices $C_1, \dots, C_{2n}$ by $CC_1C^{-1}, \dots
CC_{2n}C^{-1}$ for every symplectic matrix $C$. Here we use the
fact that every symplectic matrix corresponds to the $C$-matrix
of some Clifford operation $Q$ (Theorem \ref{thm:cr}), and
the formulas for the $C$-matrix of the inverse and multiplication of
Clifford operators (Theorem \ref{thm:cmi}).

\section{Symplectic involutions}\label{sec:symp-inv}

\noindent
In the previous section we see that each of the symplectic
matrices $C_i$ is an involution ($C_i^2=I$). Therefore, if $C_i$
was a matrix in a field of characteristic different from $2$, $C_i$
was diagonalizable with $+1, -1$ on the diagonal. In fact, a
stronger property holds; it is proved in \cite{symp} that for any
commutative set of symplectic involutions over a field of
characteristic $\neq 2$, there exists a symplectic matrix $M$ such that
$MCM^{-1}$ is diagonal with $+1,-1$ diagonal
entries, for every matrix $C$ in the set.

Here, all the symplectic involutions are over ${\mathbf{Z}}_2$, and the above
proposition does not hold anymore. However, following almost the same
steps as in \cite{symp}, an analogous result can be proved.

\begin{theorem}\label{thm:symp-inv}
For every symplectic involution $C$ of size $2n$ over ${\mathbf{Z}}_2$
there exists a symplectic matrix $M$ such that
\begin{equation}\label{eq:niceform}
MCM^{-1}=\left( \begin{array}{cc}
  I & E \\
  0 & I
\end{array} \right),
\end{equation}
where $E$ is a symmetric matrix ($E^T=E$).
\end{theorem}

\noindent
{\bf Proof:} Let us consider the block form of $C$
\begin{equation}C=\left( \begin{array}{cc}
  A & E \\
  F & B
\end{array} \right),\end{equation}
where $A, B, E$ and $F$ are $n\times n$ matrices. $C^2=I$ and
$C^T\,P\,C=P$. Then $P\,C=C^T\,P$, or equivalently $P\,C$ is
symmetric. In other words, $B=A^T$, and $E$ and $F$ are
symmetric. Therefore,
\begin{equation}C=\left( \begin{array}{cc}
  A & E \\
  F & A^T
\end{array} \right).\end{equation}
Also, $C^2=I$ gives $AE$ and $FA$ are symmetric and $A^2+EF=I$.

Assume that $rank(E)=r$. Then there is an invertible matrix $R$
such that \begin{equation}RER^T=\left( \begin{array}{cc}
  e & 0 \\
  0 & 0
\end{array} \right),\end{equation}
where $e$ is a full-rank matrix of size $r\times r$. Since $E$ is
symmetric, $e$ is symmetric as well.

Now notice that
\begin{equation}M_1=\left( \begin{array}{cc}
  R & 0 \\
  0 & R^{-T}
\end{array} \right)\end{equation}
is symplectic and the upper-right block of $M_1CM_1^{-1}$ is equal
to $RER^{T}$. So, we may assume that
\begin{equation}E=\left( \begin{array}{cc}
  e & 0 \\
  0 & 0
\end{array} \right).\end{equation}

Let
\begin{equation}A=\left( \begin{array}{cc}
  a_1 & a_{2} \\
  a_3 & a_4
\end{array} \right),\end{equation}
where the matrices $a_1, a_2, a_3$ and $a_4$ are of sizes $r\times
r, r\times (n-r), (n-r)\times r$ and $(n-r)\times (n-r)$,
respectively. $AE$ is symmetric and $e$ is invertible; then
$a_3=0$ and $a_1e$ is symmetric.

Now define
\begin{equation}S=\left( \begin{array}{cc}
  e^{-1}a_1 & e^{-1}a_2 \\
  (e^{-1}a_2)^{T} & 0
\end{array} \right),\end{equation}
and \begin{equation}M_2=\left( \begin{array}{cc}
  I & 0 \\
  S & I
\end{array} \right).\end{equation}
Since $S$ is symmetric, $M_2$ is symplectic. Also, the upper-left
block of $M_2CM_2^{-1}$ is equal to
\begin{equation}\left( \begin{array}{cc}
  0 & 0 \\
  0 & a_4
\end{array} \right).\end{equation}
Hence, we can assume that $a_1$ and $a_2$ are zero as well. (Note that in $M_2CM_2^{-1},$ $E$ remains unchanged.)

Let \begin{equation}F=\left( \begin{array}{cc}
  f_{1} & f_{2} \\
  f_{3} & f_{4}
\end{array} \right).\end{equation}
using $A^2+EF=I$ we get $a_4^2=I$, $f_1=e^{-1}$, $f_2=0$, and
$f_3=0$ since $F$ is symmetric. Then
\begin{equation}C=\left( \begin{array}{cccc}
  0 & 0 & e & 0 \\
  0 & a_{4} & 0 & 0 \\
  e^{-1} & 0 & 0 & 0 \\
  0 & f_4 & 0 & a_4^{T}
\end{array} \right).\end{equation}

Now, observe that the map
\begin{equation}\Phi(X, Y) = \left(
                 \begin{array}{cccc}
                   u_1 & 0 & u_2 & 0 \\
                   0 & v_1 & 0 & v_2 \\
                   u_3 & 0 & u_4 & 0 \\
                   0 & v_3 & 0 & v_4 \\
                 \end{array}
               \right),
\end{equation}
where
\begin{equation} X=\left(
     \begin{array}{cc}
       u_1 & u_2 \\
       u_3 & u_4 \\
     \end{array}
   \right), \hspace{.5in} Y= \left(
     \begin{array}{cc}
       v_1 & v_2 \\
       v_3 & v_4 \\
     \end{array}
   \right),
\end{equation}
preserves multiplication: $\Phi(X_1X_2, Y_1Y_2)=\Phi(X_1, Y_1)\Phi(X_2, Y_2)$.
Hence, according to the above block form of $C$, it is sufficient to prove the
theorem for the special cases
\begin{equation}C=\left( \begin{array}{cc}
   0 & e \\
  e^{-1} & 0
\end{array} \right), \hspace{.5cm} \ C=\left( \begin{array}{cc}
  a & 0 \\
  f & a^{T}
\end{array} \right).\end{equation}

In the first case if we let
\begin{equation}M=\left( \begin{array}{cc}
  I & 0 \\
  e^{-1} & I
\end{array} \right),\end{equation}
then
\begin{equation}MCM^{-1}=\left( \begin{array}{cc}
  I & e \\
  0 & I
\end{array} \right),\end{equation}
and the theorem holds.

In the second case, notice that
\begin{equation}\left( \begin{array}{cc}
  0 & I \\
  I & 0
\end{array} \right)\,C\,\left( \begin{array}{cc}
  0 & I \\
  I & 0
\end{array} \right)=\left( \begin{array}{cc}
  a^T & f \\
  0 & a
\end{array} \right).\end{equation}
Then if $f\neq 0$, by the same steps as before, we can reduce $C$
to smaller matrices and use induction. Thus, we may assume that
$f=0$, and
\begin{equation}C=\left( \begin{array}{cc}
  a^T & 0 \\
  0 & a
\end{array} \right).\end{equation}

Since $C^2=I$, $a^2=I$. Hence, there exists an invertible matrix $R$ on ${\mathbf{Z}}_2$ such that
$RaR^{-1}$ is in the Jordan normal form, and
it is a block diagonal matrix, each of its blocks is
either $I_1=(1)$ or
\begin{equation}\left( \begin{array}{cc}
  1 & 1 \\
  0 & 1
\end{array} \right).\end{equation}
On the other hand, \begin{equation}M=\left( \begin{array}{cc}
  R^{-T} & 0 \\
  0 & R
\end{array} \right)\end{equation}
is symplectic and
\begin{equation}M\,C\,M^{-1}=\left( \begin{array}{cc}
  R^{-T}a^{T}R^{T} & 0 \\
  0 & RaR^{-1}
\end{array} \right).\end{equation}
Therefore, if we prove the theorem for the two cases $a=(1)$
and
\begin{equation}a=\left( \begin{array}{cc}
  1 & 1 \\
  0 & 1
\end{array} \right),\end{equation} we are done.

In the first case there is nothing to prove, and in the second
case we have
\begin{equation}
\left( \begin{array}{cccc}
  0 & 0 & 1 & 0 \\
  0 & 1 & 0 & 0 \\
  1 & 0 & 0 & 0 \\
  0 & 0 & 0 & 1
\end{array} \right) \left( \begin{array}{cccc}
  1 & 0 & 0 & 0 \\
  1 & 1 & 0 & 0 \\
  0 & 0 & 1 & 1 \\
  0 & 0 & 0 & 1
\end{array} \right) \left( \begin{array}{cccc}
  0 & 0 & 1 & 0 \\
  0 & 1 & 0 & 0 \\
  1 & 0 & 0 & 0 \\
  0 & 0 & 0 & 1
\end{array} \right) = \left( \begin{array}{cccc}
  1 & 0 & 0 & 1 \\
  0 & 1 & 1 & 0 \\
  0 & 0 & 1 & 0 \\
  0 & 0 & 0 & 1
\end{array} \right).
\end{equation}
$\Box$

Recall that if the characteristic of the field is not $2$ we can replace
the matrix in Eq. (\ref{eq:niceform}) by a diagonal one. Also, a commutative set of symplectic involutions
over such a field can be simultaneously transformed to a set of
diagonal matrices under a symplectic change of basis
\cite{symp}. Comparing to this result and based on Theorem \ref{thm:symp-inv}, one may expect that
on a field of characteristic $2$, we can transform a commutative set of symplectic involutions to
matrices of the form of Eq. (\ref{eq:niceform}). However, this is not
the case; for a counterexample consider the following matrices:

\begin{equation}C_1=\left( \begin{array}{cccc}
  1 & 0 & 0 & 1 \\
  0 & 1 & 1 & 0 \\
  0 & 0 & 1 & 0 \\
  0 & 0 & 0 & 1
\end{array} \right), \hspace{.5in} C_2=\left( \begin{array}{cccc}
  1 & 1 & 0 & 0 \\
  0 & 1 & 0 & 0 \\
  0 & 0 & 1 & 0 \\
  0 & 0 & 1 & 1
\end{array} \right).\end{equation}
Both $C_1$ and $C_2$ are symplectic, $C_1^2=C_2^2=I$, and
$C_1C_2=C_2C_1$. If $C_1$ and $C_2$ could be
transformed to the form of Eq. (\ref{eq:niceform}), then
$(I+C_1)(I+C_2)=0$, which does not hold.

\begin{theorem}\label{thm:set-symp-inv}
Let ${\mathcal{M}}$ be a commutative set of symplectic involutions of
size $2n\times 2n$ over ${\mathbf{Z}}_2$. Then there exists a
symplectic matrix $M$ such that for every $C\in {\mathcal{M}}$,
$MCM^{-1}$ is of the form
\begin{equation}\label{eq:form}
MCM^{-1}=\left( \begin{array}{cccc}
  A & E \\
  0 & A^T
\end{array} \right),
\end{equation}
where $A$ and $E$ are $n\times n$ matrices.

\end{theorem}

\noindent
{\bf Proof:} We prove the theorem by induction on $n$. If ${\mathcal{M}}$ contains only the identity
matrix, then there is nothing to prove. So, let $C\in {\mathcal{M}}$ be
unequal to identity. According to Theorem \ref{thm:symp-inv}, we may assume
\begin{equation}C=\left( \begin{array}{cc}
  I & E \\
  0 & I
\end{array} \right).\end{equation}
Also, as in the proof of Theorem \ref{thm:symp-inv}, we may assume
\begin{equation}E=\left( \begin{array}{cc}
  e & 0 \\
  0 & 0
\end{array} \right),\end{equation}
where $e$ is a full-rank symmetric matrix of size $r\times r$.

Let $C'\in {\mathcal{M}}$ be different from $C$. If we consider the
block form of $C'$
\begin{equation}C'=\left( \begin{array}{cc}
  A' & E' \\
  F' & B'
\end{array} \right),\end{equation}
then $B'^T=A'$, and $E',F'$ are symmetric (because $C'$ is symplectic and $C'^2=I$.) Thus, we may assume
\begin{equation}C'=\left( \begin{array}{cccc}
  a_{1} & a_{2} & e_{1} &  e_{2} \\
  a_{3} & a_{4} & e_{2}^T & e_{4} \\
  f_{1} & f_{2} & a_{1}^T & a_{3}^T \\
  f_{2}^T & f_{4} & a_{2}^T & a_{4}^T
\end{array} \right).\end{equation}

Now, writing the constraint $CC'=C'C$, and using the fact that $e$
is full-rank, we conclude that $a_3=0$, $f_1=0$ and $f_2=0$.
Hence, every matrix in ${\mathcal{M}}$ is of the form
\begin{equation}\label{eq:c'}
C'=\left( \begin{array}{cccc}
  a_{1} & a_{2} & e_{1} &  e_{2} \\
  0     & a_{4} & e_{2}^T & e_{4} \\
  0     & 0   & a_{1}^T & 0       \\
  0     & f_{4} & a_{2}^T & a_{4}^T
\end{array} \right).
\end{equation}
Therefore, if $r=n$, which covers the base case $n=1$, we are done. So, assume that $n>r\geq 1$.

Suppose we map such a matrix $C'$ to its sub-matrix
\begin{equation}D'=\left( \begin{array}{cccc}
  a_4 & e_4 \\
  f_4 & a_4^T
\end{array} \right).\end{equation}
This map over matrices in the form of Eq. (\ref{eq:c'}) preserves addition and multiplication.
Therefore, all matrices $D'$ are symplectic involutions and commute. Thus, by induction we may assume $f_4=0$, and we are done. $\Box$

The following corollary is a conclusion of the above theorem and
the argument at the end of Section 3.

\begin{corollary}\label{cor:symp} To prove the claim of Theorem \ref{thm:main}, it
is sufficient to consider the case that the group ${\mathcal{G}}$ is generated by Clifford operators $Q_1, \dots, Q_{2n}$ which are represented by pairs $(C_1, h_1), \dots, (C_{2n}, h_{2n})$ such that
\begin{equation}\label{eq:nice-ci}
C_i=\left( \begin{array}{cccc}
  A_i & E_i \\
   0  & A_i^T
\end{array} \right),
\end{equation}
where $A_i^2=I$, and $E_i$ and $A_iE_i$ are symmetric. $\Box$
\end{corollary}

\newpage

\section{Clifford operators as linear transformations}\label{sec:other}

\noindent
In Section 2, we present the characterization of
Clifford operators based on how they transform Pauli matrices
under conjugation. In this section, we describe these operators as linear transformations.

In Appendix A, a Clifford operator $Q$, given by a $C$-matrix and an $h$-vector, is represented as a linear combination of Pauli matrices, and the coefficients of this expansion are extracted. These results are very general; however, according to Corollary \ref{cor:symp} for the purpose of proving our main theorem, we may assume that the matrix $C$ is in the form of Eq. (\ref{eq:nice-ci}). In the following, we show that in this special case, $Q$ is a permutation times a diagonal matrix. This result is used in Section 6 in order to find a formula which indicates whether two Clifford operators commute or anti-commute ($QQ'=Q'Q$ or $QQ'=-Q'Q$).

Assume that $Q$ is a Clifford gate, represented by the pair
$(C, h)$, where $Q^2=I$ and
\begin{equation}\label{eq:35}
C=\left( \begin{array}{cccc}
  A & E \\
  0 & A^T
\end{array} \right), \hspace{1in} h=\left( \begin{array}{c}
  f \\
  g
\end{array} \right).
\end{equation}
In this special case we can find a simpler expression for Theorem
\ref{thm:cr}. In fact,
\begin{equation} C^TJC = \left( \begin{array}{cc}
  0 & 0 \\
  0 & AE
\end{array} \right), \end{equation}
and
\begin{equation}d=diag (C^TJC)=\left( \begin{array}{c}
  0 \\
  d_0
\end{array} \right),\end{equation}
where $d_0=diag (AE)$. Thus,
\begin{equation}\label{eq:csf}
Q\tau_aQ = (i)^{d_0^T a_2} (-1)^{h^Ta+a_2^T lows(AE+d_0d_0^T)a_2}\,
\tau_{Ca},
\end{equation}
where
\begin{equation}a=\left( \begin{array}{c}
  a_{1} \\
  a_{2}
\end{array} \right).\end{equation}
Also, since $Q^2=I$ by Eq. (\ref{eq:pauli'2}),
\begin{equation}\label{eq:atf}
A^Tf=f.
\end{equation}

Let $\{\,\vert x\rangle:\, x\in {\mathbf{Z}}^{n}_2 \}$ be the
standard basis for the Hilbert space of $n$ qubits. Then we have
\begin{equation}\label{eq:cpm1}
\tau_{a}\vert x\rangle = (-1)^{a_1^T(x+a_2)}\vert
x+a_2\rangle.
\end{equation}
Let $a_2=0$. By Eq. (\ref{eq:csf})
\begin{eqnarray*}
\tau_{a} Q\, \vert x\rangle  =  Q\,(Q\tau_a Q)\,\vert x\rangle =
Q  (-1)^{f^Ta_1}\tau_{Ca} \,\vert x\rangle,
\end{eqnarray*}
and by Eq. (\ref{eq:cpm1})
\begin{equation}\tau_{a} Q\, \vert x\rangle = (-1)^{f^Ta_1}(-1)^{a_1^TA^Tx}Q\,\vert x\rangle
= (-1)^{a_1^T(f+A^Tx)}Q\, \vert x\rangle.\end{equation}
As a result, $Q\vert x\rangle
$ is the simultaneous eigenvector of all $\tau_a$, where $a_2=0$, with
eigenvalue $(-1)^{a_1^T(f+A^Tx)}$. Therefore, there exists
$\lambda_x$ such that
\begin{equation}\label{eq:lambdax}
Q\, \vert x\rangle = \lambda_x\, \vert f+A^Tx\rangle.
\end{equation}
This equation shows that the action of $Q$ on standard basis
vectors is the same as a permutation with some phases $\lambda_x$. So,
if we could compute these phases, then we had a complete
characterization of $Q$ as a linear operator.

Let $b$ be such that $b_1=0$. By Eq. (\ref{eq:lambdax}),
\begin{equation}\tau_bQ\,\vert 0\rangle = \lambda_0 \tau_b\vert f\rangle= \lambda_0\vert f+b_2\rangle.\end{equation}
Hence,
\begin{equation}\lambda_0Q\, \vert f+b_2\rangle = Q\tau_bQ\, \vert 0\rangle.\end{equation}
Equivalently,
\begin{eqnarray}\lambda_0\lambda_{f+b_2} \vert f+A^T(f+b_2)\rangle
& = & (i)^{d_0^T b_2} (-1)^{g^Tb_2+ b_2^T
lows(AE+d_0d_0^T)b_2}\,\tau_{Cb}\,\vert 0\rangle \nonumber\\
 & = & (i)^{d_0^T b_2} (-1)^{g^Tb_2+ b_2^T
lows(AE+d_0d_0^T)b_2}\, (-1)^{b_2^T AE b_2}  \,\vert A^Tb_2\rangle \nonumber \\
 & = &(i)^{d_0^Tb_2} (-1)^{d_0^Tb_2+g^Tb_2+ b_2^T
lows(AE+d_0d_0^T)b_2}\,\vert A^Tb_2\rangle,
\end{eqnarray}
where in the last equation we use $b_2^T AE b_2 = d_0^Tb_2$.
Therefore, by $A^Tf=f$ we obtain
\begin{equation}\label{eq:lambda0x}
\lambda_0\lambda_{f+y} = (i)^{d^T_0y}(-1)^{d_0^T y + g^Ty + y^T lows(AE +
d_0d_0^T)y},
\end{equation}
which determines entries of $Q$ up to an overall sign.

\begin{theorem}\label{thm:matrix-summary}
Let $Q$ be a Clifford operator, represented by $C, h$ which are
given by Eq. (\ref{eq:35}), and let $Q^2=I$. Then, the action of $Q$
on standard basis vectors is described by Eq. (\ref{eq:lambdax}),
where $\lambda_x$'s are phases that can be computed by Eq.
(\ref{eq:lambda0x}). $\Box$

\end{theorem}

\section{Sign formula} \label{sec:sign}

\noindent
Recall that in Eqs. (\ref{eq:pauli'1})-(\ref{eq:pauli'4}) we express equations
$Q_i^2=Q_j^2=I$ and $Q_iQ_j=\pm Q_jQ_i$ for two Clifford
operations $Q_i$ and $Q_j$, in terms of their $C$-matrices and
$h$-vectors. However, these formulas are independent of the plus or
minus sign, so we need another formula, which we call the sign
formula, to recognize two cases $Q_iQ_j=Q_jQ_i$ and $Q_iQ_j=-
Q_jQ_i$.

Suppose $Q=Q_i$ and $Q'=Q_j$ are represented by $(C=C_i, h=h_i)$
and $(C'=C_j, h'=h_j)$, respectively, and satisfy Eqs.
(\ref{eq:pauli'1})-(\ref{eq:pauli'4}). Following Corollary
\ref{cor:symp}, let us assume
\begin{equation}\label{eq:cc'}C=\left( \begin{array}{cc}
  A & E \\
   0  & A^T
\end{array} \right),
\hspace{.5in} C'=\left( \begin{array}{cc}
  A' & E' \\
   0  & A'^T
\end{array} \right),
\end{equation}
where $A^2=A'^2=I$. Also, let
\begin{equation}\label{eq:hh'} h=\left( \begin{array}{c}
  f \\
  g
\end{array} \right), \hspace{.5in} h'=\left( \begin{array}{c}
  f' \\
  g'
\end{array} \right).\end{equation}
Then by Eq. (\ref{eq:pauli'4}) we have $f+A^Tf'=f'+A'^Tf$, or
equivalently
\begin{equation}\label{eq:f-f'}
(I+A^T)f'=(I+A'^T)f.
\end{equation}

According to Eq. (\ref{eq:lambdax})
\begin{equation}Q'Q\,\vert 0\rangle = Q'\lambda_0\,\vert f\rangle = \lambda_0\lambda'_f\, \vert f'+A'^T f\rangle.\end{equation}
Similarly, $QQ'\vert 0\rangle=\lambda_{f'}\lambda'_0\, \vert f + A^Tf'\rangle$.
Therefore, we conclude that $Q'Q=-QQ'$ if and only if
$\lambda_0\lambda'_f=-\lambda'_0\lambda_{f'}$, or equivalently
\begin{equation}\label{eq:sign}(\lambda_0^{\,2})(\lambda'_0\lambda'_f)=-(\lambda'^{\,2}_0)(\lambda_0
\lambda_{f'}).
\end{equation}
Now note that using Eq. (\ref{eq:lambda0x}) we can
explicitly compute all the terms in this equation. As a result, the sign in $Q'Q=\pm
QQ'$ can be determined in terms of $C, h, C'$, and $h'$. We do not express this formula here because of its complexity; however, it is clear from Eq. (\ref{eq:sign}) that if $f=f'=0$, then $Q$ and $Q'$ commute.

\begin{lemma} \label{lem:sign} Suppose $Q$ and $Q'$ are represented by Eqs. (\ref{eq:cc'}) and (\ref{eq:hh'}), and satisfy $Q^2=Q'^2=I$ and $QQ'=\pm Q'Q$. Then, if $f=f'=0$, $QQ'=Q'Q$. $\Box$
\end{lemma}

\section{Proof of Theorem \ref{thm:main-result}} \label{sec:proof}

\noindent
Using Theorem \ref{thm:main}, let ${\mathcal{G}}$ be a subgroup of the Clifford group, isomorphic to the Pauli group.
Suppose that ${\mathcal{G}}$ is generated by $Q_1,\dots, Q_{2n}$ which are represented
by pairs \begin{equation}(C_1, h_1), \dots, (C_{2n},h_{2n}),\end{equation} respectively, and satisfy Eqs. (\ref{eq:pauli1}) and (\ref{eq:pauli2}),
and then (\ref{eq:pauli'1})-(\ref{eq:pauli'4}). By Corollary
\ref{cor:symp}, we may assume that
\begin{equation}C_i=\left( \begin{array}{cc}
  A_{i} & E_i \\
  0 & A^T_i
\end{array} \right),\hspace{1in} h_i=\left( \begin{array}{c}
  f_{i} \\
  g_{i}
\end{array} \right).\end{equation}
We prove that there exists a maximal abelian subgroup ${\mathcal{H}}$ of ${\mathcal{G}}$ such that all of matrices in ${\mathcal{H}}$ are diagonal. In that case, the linear span of ${\mathcal{H}}$ would be equal to the linear span of the group generated by all $\sigma_z$ operators (which is a maximal abelian subgroup of the Pauli group), and we are done.

Define the map $T: {\mathbf{Z}}^{2n}_2\rightarrow {\mathbf{Z}}^n_2$
that sends $x=(x_1, \dots ,x_{2n})$ to the $f$-vector of the
Clifford operator $Q^{x_1}_1  \dots Q^{x_{2n}}_{2n} $. (By
$f$-vector we mean the upper part of its $h$-vector.) $T$ is not
linear but {\it almost linear}.

\begin{lemma} \label{lem:map-t}
        \hfil

\begin{enumerate}

\item[{\rm (i)}] $Ker\, T=T^{-1}(0)$ is a linear subspace of ${\mathbf{Z}}^{2n}_2$.

\item[{\rm (ii)}] For any $x$ and $x'$, where $T(x)=T(x')$, $x+x'\in Ker\, T$.

\item[{\rm (iii)}] If $y\in Ker\, T$, then $T(x+y)=T(x)$, for any $x$.

\item [{\rm (iv)}] For every $y\in {\mathbf{Z}}^n_2$, $T^{-1}(y)$ is either empty or equal to $x+Ker\,
T$, for some $x\in  {\mathbf{Z}}^{2n}_2$.

\item[{\rm (v)}] $\vert Ker\, T\vert . \vert Im\, T\vert= 2^{2n}$

\end{enumerate}

\end{lemma}

\noindent
{\bf Proof:}
Define ${\mathcal{Q}}_x=Q^{x_1}_1\dots Q^{x_{2n}}_{2n}$ and
${\mathcal{A}}_x=A^{x_1}_1\dots A^{x_{2n}}_{2n}$.

{\rm (i)} We should show that if the $f$-vectors of
${\mathcal{Q}}_x$ and ${\mathcal{Q}}_{x'}$ are both zero, then the
$f$-vector of ${\mathcal{Q}}_{x+x'}=\pm
{\mathcal{Q}}_x{\mathcal{Q}}_{x'}$ is also zero. By Theorem
\ref{thm:cmi} the $f$-vector of ${\mathcal{Q}}_x{\mathcal{Q}}_{x'}$
is equal to $T(x)+{\mathcal{A}}^T_x T(x')=0 $.

{\rm (ii)} Let $T(x)=T(x')$. Then, again by Theorem
$\ref{thm:cmi}$ and Eq. (\ref{eq:atf})
\begin{equation}T(x+x')= T(x)+ {\mathcal{A}}^T_xT(x') = T(x)+ {\mathcal{A}}^T_xT(x) = 0.\end{equation}
Thus, $x+x'$ is in $Ker\, T$.

{\rm (iii)} $T(x+y)=T(x)+{\mathcal{A}}^T_x T(y)=T(x)$.

{\rm (iv)} and {\rm (v)} are clear from {\rm (ii)} and {\rm (iii)}. $\Box$

\begin{lemma}\label{lem:ker-t} $dim\, Ker\, T=n$, and $T$ is
surjective.
\end{lemma}

\noindent
{\bf Proof:} By the previous lemma, $\vert
Ker\, T\vert . \vert Im\, T\vert= 2^{2n}$, so if we prove $dim\,
Ker\, T=n$, then $T$ would be surjective automatically.

$\vert Im\, T\vert\leq 2^n$, then $\vert Ker\,
T\vert \geq 2^n$, or equivalently $\dim Ker\, T =r\geq n$.

Let $x, x'\in Ker\, T$, so by definition, the $f$-vector of
${\mathcal{Q}}_x$ and ${\mathcal{Q}}_{x'}$ are both zero. Thus, by Lemma \ref{lem:sign}, ${\mathcal{Q}}_x$ and
${\mathcal{Q}}_{x'}$ commute. Therefore, since by Lemma
\ref{lem:map-t}, $Ker\, T$ is a linear subspace, ${\mathcal{H}}=\{ \pm{\mathcal{Q}}_x, \pm i{\mathcal{Q}}_x:
\, x\in KerT\}$ is an abelian subgroup of ${\mathcal{G}}$. On the other hand, every
maximal abelian subgroup of the Pauli group is of size $4\times 2^n$. (The factor $4$ is duo to the phases $\pm 1$ and $\pm i$.) Hence, $4\times 2^r=\vert \{ \pm {\mathcal{Q}}_x, \pm i{\mathcal{Q}}_x: \, x\in KerT\}\vert \leq 4\times 2^n$. As a result, $dim\, Ker\, T=r=n$. $\Box$

This lemma and its proof show that ${\mathcal{H}}$ is a maximal abelian subgroup of ${\mathcal{G}}$.

\begin{lemma} ${\mathcal{A}}_x=I$, for any $x\in Ker\, T$.
\end{lemma}

\noindent
{\bf Proof:} By Eq. (\ref{eq:f-f'}), if $x\in Ker\,T$, we have
\begin{equation}(I+{\mathcal{A}}^T_{x})T(y)=(I+{\mathcal{A}}^T_y)T(x)=0,\end{equation}
for every $y$. On the other hand, by Lemma \ref{lem:ker-t}, $T$ is surjective; thus,
$(I+{\mathcal{A}}^T_{x})y=0$, for every vector $y$. Equivalently,
${\mathcal{A}}_{x}=I$. $\Box$

Now, we are ready to finish the proof. ${\mathcal{H}}$ is a maximal abelian group of ${\mathcal{G}}$ which is a group isomorphic to the Pauli group. Therefore, if we show that all elements of ${\mathcal{H}}$ are diagonal, the linear span of ${\mathcal{H}}$ is equal to the linear span of the maximal abelian subgroup of the Pauli group generated by $\sigma_z$ gates, and then we are done.

Let $x\in Ker\, T$; we prove that ${\mathcal{Q}}_x$ is diagonal.  Using Eq. (\ref{eq:lambdax}), since the $f$-vector of ${\mathcal{Q}}_x$ is zero, and ${\mathcal{A}}_x=I$, ${\mathcal{Q}}_x\vert y\rangle =\lambda_y \vert y\rangle$, for some $\lambda_y$. This equality means that ${\mathcal{Q}}_x$ is diagonal in the
standard basis. We are done.

\section{Conclusion}

\noindent
In this paper we develop some techniques to characterize
${\mathcal{C}}_3$ gates based on the subgroups of the Clifford group
isomorphic to the Pauli group. We prove that any such group, after
conjugation by a Clifford operation, contains a maximal abelian
subgroup, all of whose elements are diagonal. This result proves
the conjecture that any ${\mathcal{C}}_3$ gate is a generalized
semi-Clifford gate. Using Proposition 2 of \cite{bei}, we conclude
that any ${\mathcal{C}}_3$ gate is of the form $Q\Pi\Lambda Q'$,
where $\Pi$ is a permutation, $\Lambda$ is diagonal, and $Q, Q'$
are Clifford operations. To obtain a deeper understanding of
${\mathcal{C}}_3$ gates we should characterize all
of these groups (subgroups of Clifford group isomorphic to Pauli
group). Such a characterization leads us to a better
understanding of ${\mathcal{C}}_3$, and then ${\mathcal{C}}_k$,
$k\geq 4$.\\

\noindent{\bf Acknowledgements.}
Authors are thankful to
Carlos Mochon, Daniel Gottesman, and Bei Zeng for providing the
counterexample of Conjecture \ref{conj1}. They are also grateful to unknown referees for their comments which improved the presentation of the results.

\appendix

\vspace{.35in}

\begin{center} {\large \bf Appendix A}
\end{center}

\noindent
This appendix contains some results regarding the coefficients of the linear expansion of a Clifford operator in terms of Pauli matrices.

Let $Q$ be an arbitrary Clifford operation represented by the pair
$(C,h)$. Since Pauli operators consist a basis for the linear space of
matrices, there are complex numbers $r_a$, $a\in {\mathbf{Z}}_2^{2n}$, such that
\begin{equation}\label{eq:lpm}
Q=\sum_a r_a\, (i^{a^TJa}\,\tau_a).
\end{equation}
Using Eqs. (\ref{eq:conj1})-(\ref{eq:conj2}),
for every $b\in {\mathbf{Z}}_2^{2n}$ we have
\begin{equation}Q\,\tau_b\, Q^\dagger=(i)^{d^T b}(-1)^{h^T b+b^T lows(C^TJC+dd^T)b}\,\tau_{Cb},\end{equation}
where $d$ is defined by Eq. (\ref{eq:d}), and then
\begin{equation}Q\tau_b = (i)^{d^T b}(-1)^{h^T b+b^T lows(C^TJC+dd^T)b}\,\tau_{Cb}Q.\end{equation}
Therefore, replacing $Q$ by Eq. (\ref{eq:lpm}), we get
$$\sum_a r_a (i)^{a^TJ\,a} (-1)^{b^TJ\,a} \tau_{a+b}\hspace{3.4in}$$
\vspace{-.19in}
\begin{equation}\hspace{.8in}=(i)^{d^T b}(-1)^{h^T b+b^T lows(C^TJC+dd^T)b}
\sum_{a'} r_{a'}(i)^{a'^TJ\,a'} (-1)^{a'^TJ\,Cb}\tau_{a'+Cb}.\end{equation}
Equivalently, if $a'=a+b+Cb$, then
\begin{equation}\label{eq:a'} r_a\, (i)^{a^TJ\,a} (-1)^{b^TJ\,a}=r_{a'}\, (i)^{d^T
b}(i)^{a'^TJ\,a'}(-1)^{h^T b+b^T lows(C^TJC+dd^T)b}
(-1)^{a'^TJ\,Cb}.
\end{equation}

Suppose $b$ is an eigenvector of $C$ with eigenvalue one ($Cb=b$).
Then $a'=a$, and by the above equation if $r_a\neq 0$
\begin{equation}\label{eq:a}(-1)^{b^TJ\,a}=(i)^{d^Tb}(-1)^{h^T b+b^T
lows(C^TJC+dd^T)b}(-1)^{a^TJ\,b}.
\end{equation}
Thus, $d^Tb$ must be zero.

To see this fact more explicitly, consider the block form of $C$
\begin{equation}C=\left( \begin{array}{cc}
  A & E \\
  F & B
\end{array}\right).\end{equation}
Since $C$ is symplectic, $B^TA+E^TF=I$ and $A^TF, E^TB$ are
symmetric. Also, $Cb=b$ is equivalent to
\begin{equation}\label{eq:eig}
Ax+Ey=x,\hspace{1in} Fx+By=y.
\end{equation}
where
\begin{equation}b=\left( \begin{array}{c}
  x \\
  y
\end{array}\right).\end{equation}

By the definition of $d$,
\begin{eqnarray}
d  =  diag(C^TJC)& = & diag
\left( \begin{array}{cc}
  A^TF & A^TB \\
  E^TF & E^TB
\end{array}\right) \nonumber\\ \label{eq:symmetry}
 & = & diag \left( \begin{array}{cc}
  F^TA & F^TE \\
  E^TF & E^TB
\end{array}\right)\nonumber\\
 & = & diag \left[\left( \begin{array}{cc}
  F^T &  0 \\
   0 & E^T
\end{array}\right)C\right].
\end{eqnarray}
Since the last matrix is symmetric, and the operations are on a
field of characteristic $2$, we have
\begin{equation}d^Tb= b^T \left( \begin{array}{cc}
  F^T &  0 \\
   0 & E^T
\end{array}\right)C\, b=b^T \left( \begin{array}{cc}
  F^T &  0 \\
   0 & E^T
\end{array}\right) b= x^TF^Tx+y^TE^Ty.\end{equation}
Then we should show that $x^TF^Tx+y^TE^Ty=0$. In fact, we can prove a
stronger equality:
\begin{equation}b'^T \left( \begin{array}{cc}
  F^T &  0 \\
   0 & E^T
\end{array}\right) b=x'^TF^Tx+y'^TE^Ty=0,$$ where
$$b'=\left( \begin{array}{c}
  x' \\
  y'
\end{array}\right)\end{equation}
is another eigenvector of $C$ with eigenvalue one.

Using $B^TA+E^TF=I$, we have $y'^TB^TAx+y'^TE^TFx=y'^Tx$, and by
Eq. (\ref{eq:eig}) we conclude that
\begin{eqnarray}
0 &=& y'^TB^TAx+y'^TE^TFx+y'^Tx \nonumber\\
  &=& y'^TB^T(x+Ey)+y'^TE^T(y+By)+y'^Tx \nonumber\\
  &=& y'^TB^Tx+y'^TE^Ty+y'^Tx \nonumber\\
  &=& y'^TE^Ty+ (y'^T+x'^TF^T)x+y'^Tx \nonumber\\
  &=& y'^TE^Ty+ x'^TF^Tx,
\end{eqnarray}
where in the third line we use $B^TE=E^TB$.

\begin{lemma}\label{lem:linear} The map $S(b)=b^T lows(C^TJC+dd^T)b$ that is
defined on the set of eigenvectors of $C$ with eigenvalue one is a linear
map. As a result, there exists a vector $\alpha$ such that $S(b)=\alpha^T
b$.
\end{lemma}

\noindent{\bf Proof:} It is sufficient to show that $b^T lows(C^TJC+dd^T)b'+b'^T
lows(C^TJC+dd^T)b=0$ for two eigenvectors $b$ and $b'$ with eigenvalue one.
$$b^T lows(C^TJC+dd^T)b'+b'^T lows(C^TJC+dd^T)b \hspace{2in}$$
\vspace{-.35in}
\begin{eqnarray}
 & = & b^T lows(C^TJC+dd^T)b'+b^T lows(C^TJC+dd^T)^Tb' \nonumber\\
 & = & b^T\left[lows(C^TJC+dd^T)+lows(C^TJC+dd^T)^T\right]b' \nonumber\\
 & = & b^T\left[ \left( \begin{array}{cc}
   F^T & 0 \\
    0 & E^T \
 \end{array}\right) C+dd^T  \, \right]b' \nonumber\\
 & = & b^T \left( \begin{array}{cc}
   F^T & 0 \\
    0 & E^T \
 \end{array}\right) b'+ b^Tdd^T b' \nonumber\\
 & = & 0,
\end{eqnarray}
where in the fourth line we used the same idea as in Eq. (\ref{eq:symmetry}), and the fact that $d=diag\, (dd^T)$ which gives \begin{equation}diag\, \left[\left( \begin{array}{cc}
   F^T & 0 \\
    0 & E^T \
 \end{array}\right) C+dd^T \right]=0.\end{equation}
$\Box$

Now, let us return to Eq. (\ref{eq:a}). If $r_a\neq 0$, then for any $b$ such that $Cb=b$ we have
\begin{equation}(-1)^{b^TJ\,a}=(-1)^{h^T b+b^T lows(C^TJC+dd^T)b}(-1)^{a^TJ\,b},\end{equation}
or equivalently
\begin{equation}a^TP\,b+ h^T b+b^T lows(C^TJC+dd^T)b=0.\end{equation}
Using Lemma \ref{lem:linear}, we can write this equality as
\begin{equation}\label{eq:nz}
(Pa+ h+\alpha )^T b=0.
\end{equation}

Suppose $dim\, Ker(I+C)=s$, i.e., there are $s$ independent
eigenvectors of $C$ with eigenvalue one. Thus, by Eq. (\ref{eq:nz}),
there are $s$ independent linear constraints on the vectors $a$ for which $r_a\neq 0$. Therefore, there are at most $2^{2n-s}$ vectors
$a$ such that $r_a\neq 0$. On the other hand, if $r_a$ is non-zero,
then by Eq. (\ref{eq:a'}), $r_{a+e}\neq 0$, for every vector $e$
in the image of $I+C$. Moreover, $dim\, Im(I+C)=2n-s$, and then
the number of such vectors $e$ is equal to $2^{2n-s}$. This means that all of vectors $a'$ for which
$r_{a'}$ is non-zero, are of the form $a'=a+e$ for some $e\in Im(I+C)$.

As a summary, we have the following theorem.

\begin{theorem}\label{thm:coeff} Let $Q$ be a Clifford operation represented by the pair
$(C, h)$, and let $r_a$, $a\in {\mathbf{Z}}_2^{2n}$, be the coefficients of $Q$ as a linear
combination of Pauli matrices as in Eq. (\ref{eq:lpm}). Also, let
$\alpha$ be the vector defined in Lemma \ref{lem:linear}. Then,
$r_a$ is non-zero for $a=P(h+\alpha)$. Moreover, every $a'$ where
$r_{a'}\neq 0$, is of the form $a'=P(h+\alpha)+b+Cb$ for some $b$.
In this case, $r_a$ and $r_{a'}$ are related by Eq.
(\ref{eq:a'}). In particular, $\vert r_a\vert=\vert r_{a'}\vert$. $\Box$
\end{theorem}

\end{document}